# Targetability of chaotic sets with small parameter perturbations

Xiao-Song Yang

Department of Mathematics,

Huazhong University of Science and Technology.

Wuhan, 430074, China

*Abstract*

In this paper targetability of chaotic sets with small controls is discussed by virtue of some results of geometric control theory. It is proved that given a chaotic set $\Lambda$, it is possible to steer a orbit in $\Lambda$ to every final state in some neighborhood of the chaotic set by suitable small perturbations to the parameters of the system under the Lie rank condition.

## 1 Introduction

Over the years, control of chaotic dynamics has evolved as one of the central issues in applied nonlinear science since the seminal article by Ott, Grebogi, and Yorke was published in 1990. In their article it was demonstrated that small time-dependent changes in the control parameters of a nonlinear system can turn a previously chaotic trajectory into a stable periodic motion. Nowadays the notion of chaos control has been extended to a much wider class of problems involving stabilization of unstable periodic or stationary states and targeting trajectories to some final states in nonlinear dynamic systems. Within the last few decades a considerable progress has been made in this field, in particular with respect to extending the methods of chaos control to spatiotemporal patterns, extending the methods of control of deterministic dynamic systems to stochastic and noise-mediated systems, and applications to various areas, e.g., biological, medical, technological systems.[1-3, 6-14] （see [13] and references therein）. In addition a lot of traditional control technique have been utilized to stabilize a vast of specific physics and engineering chaotic systems [1,2,6] （see [13] and references therein）.

The basic idea of chaos control is exploiting the key dynamic characteristics just presented in a system to control the system so that the controlled system has desired dynamic behavior [7]. The remarkable feature of chaos control is that the control is tiny and is applied locally in state space and intermittently or occasionally in time. Thus the fundamental idea of chaos control is applied like this: the system flexibility is paramount and opportunistically exploited so that the controls (perturbations ) do not significantly change the system dynamics, but just enable the system's intrinsic dynamics to accomplish the desired control task. Furthermore, to fully exploit the flexibility presented on chaotic systems, the controller that implements the chaos control strategy must preserve the original dynamics at most of time. In some sense, the control of chaos follows the way that Nature implements its control strategy to accomplish its goals.

There are two main lines in chaos control studies:

1 Various (open or feedback) control approaches based on characteristic of chaotic sets are pursed to suppress chaotic behavior by means of stabilizing periodic orbits or equilibrium points

[7]

2 Besides, finding some suitable small perturbations, i.e., small control strategies applied to some control parameter in order to direct trajectories to some desired final state has also received intensive investigations [7]. This problem of guiding trajectories in chaotic systems to some final state is called targeting in the chaos control literature [7, 12].

However, a basic question relevant to chaos control, which is of paramount significance from theoretical point of view, has not been touched upon systematically to the best knowledge of the author: the targetability of chaotic sets, or in terms of control theory, controllability of chaotic set by small system perturbations (control).

As for this problem it should be emphasized that in practice one can not control a systems by arbitrary perturbations to the chaotic system, the perturbations to the system must obey some physical constraints, or only through some admissible parameters.

Therefore one has to face the following general problem: can one tune some parameters of a chaotic system slightly to influence its dynamic behavior as one likes? More precisely the question concerning targetability of chaotic sets is formulated as follows:

Consider the following system

$$\dot{x} = f(x,u), \quad x \in R^n, \quad u \in U \subset R^m,$$

Where $f$ is smooth and $U$ is a compact set with $0 \in \text{int}\, U$.

Let $\Lambda \subset R^n$ be an invariant subset of the following system

$$\dot{x} = f_0(x) = f(x,0)$$

Specifically suppose that $\Lambda$ is a chaotic set (attractor).

*Given a chaotic set $\Lambda$, is it possible to steer a orbit in $\Lambda$ to every final state in some neighborhood of the chaotic set by suitable small perturbations to the parameters of the system?*

Closely associated with the above question is the following question that is of interest of its own.

Let $\gamma \subset \Lambda$ be a periodic orbit (trajectory), can we adjust the parameter such that this periodic orbit can be steered to nearby (periodic) orbit? Or equivalently, can we adjust parameter slightly so that we can steer the nearby orbits to the given periodic orbit $\gamma$?

Clearly the affirmative answer to this question is of remarkable significance to stabilization of unstable periodic orbits in chaotic systems.

In this paper we will discuss these issues by virtue of some results of geometric control theory.

## 2 Preliminaries: geometric control theory and Chaotic sets

For the control system

$$\dot{x} = f(x,u), \quad x \in R^n, \quad u \in U \subset R^m.$$

we recall some preliminaries and results from geometric control theory for details see any text book on this subject. First we recall the concept of reachable set.

A point $\bar{x}$ is said to be accessible from a given point $x_0$ if the exists a $T > 0$, and an admissible (Lebesgue integrable) control function $u(t):[0,T] \to U$, such that the solution $\phi(t,x_0,u)$ to the control system

$$\dot{x} = f(x,u(t))$$

with initial condition $\phi(0,x_0,u) = x_0$ satisfies $\phi(T,x_0,u) = \bar{x}$. For a given point $x_0$, its reachable set $R(x_0)$ is defined to be the all points that can be accessible from $x_0$ in finite time. Specifically one can also define a reachable set $R(x_0,T)$ to be the set of the all points that can be accessible from $x_0$ at time $T$. It is easy to see that $R(x_0,T) \subset R(x_0)$.

Let $\phi(x_0,t)$ be the solution to the following system with $u = 0$

$$\Sigma_0: \quad \dot{x} = f_0(x) = f(x,0)$$

**Definition 2.1** The systems is said to be locally controllable along the orbit $\phi(x_0,t)$ of $\Sigma_0$, if for each $t > 0$, $\phi(x_0,t) \in \text{int } R(x_0,t)$, where $\text{int } R(x_0,t)$ means the interior of $R(x_0,t)$.

Now let us recall the concept of Lie bracket of vector fields. Let $f$ and $g$ be two smooth vector fields on a differentiable manifold $M$. The Lie bracket $[f,g]$ can be defined as follows: For every smooth function $V$ defined on $M$,

$$[f,g]V = f(gV) - g(fV),$$

where $fV$ is the Lie derivative of $V$ along the vector field $f$.

In local coordinates the Lie bracket of $f$ and $g$ is given in the following relation

$$[f,g] = \frac{\partial g}{\partial x} f - \frac{\partial f}{\partial x} g.$$

Now consider control affine systems of the following type

$$\Sigma: \quad \dot{x} = f_0(x) + u_1 f_1(x) + \cdots + u_m f_m(x),$$

and the uncontrolled system

$$\Sigma_0: \quad \dot{x} = f_0(x).$$

Let

$$S(x_0) = span\{ad_{f_0}^k f_i : k \geq 0, i = 1, \cdots, m\}\big|_{x=x_0}$$

where

$$ad_{f_0}^0 f_i = f_i, ad_{f_0}^1 f_i = [f_0, f_i], \cdots, ad_{f_0}^k f_i = [f_0, ad_{f_0}^{k-1} f_i], \cdots.$$

For an orbit $\phi(x_0, t)$ of the system $\Sigma_0$ with initial condition $\phi(x_0, 0) = x_0$, a well known result in geometric control theory is the following proposition

**Proposition 2.1**[4,5] If there exists $x_0 \in R^n$ such that the following Lie rank condition holds

$$\dim S(x_0) = n,$$

then $\Sigma$ is locally controllable along the orbit $\phi(x_0, t)$. The equality in this proposition is called rank condition of $S$ at $x_0$.

Reversing the time variable, from **Proposition 2.1** one can easily prove the following fact.

**Proposition 2.2** Suppose that the control set $U \subset R^m$ is symmetric, i.e., $u \in U \Rightarrow -u \in U$, if there exists $x_0 \in R^n$ such that

$$\dim S(x_0) = n,$$

Then for each $t > 0$,

$$\phi(x_0, -t) \in int\, R(x_0, -t).$$

This implies that for every point $x \in int\, R(x_0, -t)$ one can steer $x$ to $x_0$.

Let $M$ be a smooth manifold, $\phi(t, *)$ be smooth flow on $M$

$$\phi(t, *) : M \to M,.\ t \in R$$

which is generated by the following differential equation

$$\dot{x} = f(x),\ x \in M$$

**Definition 2.3** An invariant compact set $\Lambda \subset M$ is a hyperbolic set for $\phi(t, *) : M \to M$ if the tangent bundle over $\Lambda$ admits a continuous decomposition

$$T_\Lambda M = E^u \oplus E^o \oplus E^s$$

where $E^o$ is one-dimensional and collinear to the flow direction, i.e., $\dfrac{d}{dt}\phi(t, *) \subset E^o$: in case that $\Lambda$ consists of equilibria, $E^o = \emptyset$.

Denote by $Per(\phi)$ the set of periodic orbits of the flow $\phi(t,*)$. In this paper we consider the chaotic set defined in the following way.

**Definition 2.4.** An invariant compact set $\Lambda \subset M$ is a chaotic set for $\phi(t,*): M \to M$ if it satisfies

(i) $\Lambda$ is transitive

(ii) $\Lambda$ is hyperbolic

(iii) $Per(\phi)$ is dense in $\Lambda$.

It is easy to see that if (i) and (ii) hold then periodic orbit set $Per(\phi)$ is dense in $\Lambda$. We include (iii) in the definition just for convenience of later arguments.

In this paper we focus our attention on the control affine system of the following form

$$\Sigma: \quad \dot{x} = f_0(x) + u_1 f_1(x) + \cdots + u_m f_m(x),$$

due to the following considerations.

Again we consider the following general control system:

$$\dot{x} = f(x,u), \quad x \in R^n \quad u \in R^m$$

where $u$ is a parameter in physical terminology or input ( control ) in control terminology.

As everyone knows in chaos control one is more interested in small variation of parameter or small input that can affect dynamics of chaotic system, thus one has the following problem.

Assume that system

$$\dot{x} = f_0(x) = f(x,0)$$

has a chaotic set somewhere in state space $R^n$. For a sufficiently small number $\varepsilon > 0$, one wants to investigate the following chaos control system

$$\Sigma_c : \dot{x} = f(x,u), \quad \|u\| < \varepsilon,$$

This implies we should consider the following control affine system if $u$ is small enough

$$\Sigma: \quad \dot{x} = f_0(x) + u_1 f_1(x) + \cdots + u_m f_m(x),$$

where $f_0(x) = f(x,0)$, $f_i(x) = \dfrac{\partial f(x,u)}{u_i}\Big|_{u=0}$, $i = 1, \ldots, m$.

Thus the fundamental question is now whether $\Sigma$ can be controlled by small inputs $u$.

Now we give the definition of targetability of an invariant set.

**Definition 2.5** An invariant compact set $\Lambda$ of the following system

$$\Sigma_0: \quad \dot{x} = f_0(x)$$

is said to be targetable, if there exists neighborhood $N_\Lambda$ of $\Lambda$, such that for any two points $q_1$ and $q_2$ contained in $N_\Lambda$, there exists a piecewise control $u(t)$ such that the corresponding control orbit $\phi(t,p,u))$ with $\phi(0,q_1,u)=q_1$ satisfies $\phi(T,q_1,u))=q_2$ for some $T>0$.

### 3 Targetbility in neighborhood of periodic orbits by small control.

By virtue of **Proposition 2.1**, we can prove the following statement.

**Proposition 3.1** Let $\gamma \subset \Lambda$ be a periodic orbit of $\Sigma_0$. Suppose that there exist a point $p \in \gamma$ such that

$$\dim S(p) = n.$$

Then there exist a neighborhood $N_\gamma$ of $\gamma$. For each point $q \in N_\gamma$, there is a piecewise constant control $u_q(t)$ such that the control orbit $x(t,p,u_q)$ with $x(0,p,u_q)=p$ satisfies $x(\widehat{T},p,u_q)=q$ for some $\widehat{T}>0$.

As a consequence, one can assert that one can steer the periodic orbit $\gamma$ to near by orbits with suitable control.

***Proof.*** Denote by $x(t,p)$ the periodic orbit $\gamma$. Let T be the period of $\gamma$ i.e., $\phi(t+T,p)=\phi(t,p)$. Then by **Proposition 2.1**, $p=\phi(T,p) \in \text{int } R(p,T)$. Thus there is an open ball $B(p)$ centered at $p$ satisfying

$$p \in B(p) \subset \text{int } R(p,T)$$

Let $K_p$ be the Poincaré section surface to the orbit $\phi(p,t)$ at the point $p$, then the set $\widetilde{K}=K_p \bigcap B(p)$ ($B(p)$ can be chosen smaller if necessary) is an open set contained in $K_p$. For every point $x \in \widetilde{K}$, it is easy to see that there is a piecewise constant control $u_x(t)$ defined on $[0,T]$ such that the control orbit $\phi(t,p,u_x)$ satisfies $\phi(0,p,u_x)=p$ and $\phi(T,p,u_x)=x)$ in view of the definition $B(p)$. Now consider the set

$$\{\phi(\widetilde{K},t), t \in [0,\widetilde{T}], \widetilde{T}>T\}$$

It is clearly contains an open tubelar neighborhood of the orbit $\gamma$. Denote it by $N_\gamma$, then for every $q \in N_\gamma$, there is point $x \in \widetilde{K}$, such that $\phi(x,\hat{t})=q$ for some $\hat{t} \in [0,\widetilde{T}]$. This implies that there is a piecewise constant control $u_q$ defined on $[0, T+\tilde{T}]$ which is $u_x$ on $[0, T]$ and is zero on $[T, T+\hat{t}]$ such that the control $u_q(t)$ can steer the point $p$ to $q$. □

Reversing the time variable, we can prove the following fact be virtue of **Proposition 2.2.**

**Proposition 3.2**: Let $\gamma \subset \Lambda$ be a periodic orbit, Suppose that there exist a point $p \in \gamma$ such that

$$\dim S(p) = n.$$

Then there exist a neighborhood $N^\gamma$ of $\gamma$. For every point $q \in N^\gamma$, there is a piecewise constant control $u(t)$ such that the control orbit $\phi(t, q, u)$ with $\phi(0, q, u) = q$ satisfies $\phi(\bar{t}, q, u) = p$, for some $\bar{t} > 0$. This is implies that one can steer nearly orbits to $\gamma(t)$ by piecewise constant control $u(t)$ in finite time.

## 4 Targetability of chaotic sets with small controls

With **Proposition 3.1** and **Proposition 3.2** we have the following theorem

**Proposition 4.1.** Let $\gamma \subset \Lambda$ be a periodic orbit of $\Sigma_0$. Suppose that there exist a point $p \in \gamma$ such that

$$\dim S(p) = n.$$

Then there is an open neighborhood $N(\gamma)$ of $\gamma$ such that for every pair points $q_1, q_2 \in N(\gamma)$, there is a piecewise constant control $u(t)$ such that the control orbit $\phi(t, q_1, u)$ with $\phi(0, q_1, u) = q_1$ satisfies

$$\phi(T, q_1, u) = q_2, \text{ for some } T > 0.$$

*Proof* Let $N^\gamma$ be the neighborhood defined in **Proposition 3.2** and $N_\gamma$ be the neighborhood considered in **Proposition 3.1**. Then $N^\gamma \cap N_\gamma$ contains an open neighborhood $N(\gamma)$ of $\gamma$ that has the following property. For any two point $q_1$ and $q_2$ contained in $N(\gamma)$, there is a piecewise constant control to steer $q_1$ to $q_2$. Since $q_1 \in N_\gamma$, there is $\bar{u}(t)$ such that the control orbit $\phi(t, q_1, \bar{u})$ with $\phi(0, q_1, \bar{u}) = q_1$ satisfies $\phi((\bar{t}, q_1, \bar{u})) = p$ for some $\bar{t} > 0$ in view Proposition 3. Since $q_2 \in N^\gamma$, there is $\hat{u}(t)$ such that the control orbit $\phi(p,\hat{u}(t))$ with $\phi(p,0) = p$ and $\phi(p,\hat{u}(\hat{t})) = q_2$ for some $\hat{t} > 0$.

Now define

$$u = \begin{cases} \bar{u}(t), & t \in [0,\bar{t}) \\ \hat{u}(t), & t \in [\bar{t},\bar{t}+\hat{t}] \end{cases}$$

It is easy to see that the corresponding control orbit $\phi(t,q_1,u)$ satisfies $\phi(0,q_1,u) = q_1$ and $\phi(\bar{t}+\hat{t},q_1,u) = q_2$. The proof is complete. $\square$

Now we prove the following theorem.

**Theorem 4.1**. Suppose $\Lambda$ is chaotic set of

$$\Sigma_0 : \quad \dot{x} = f_0(x)$$

with the following rank property

$$\dim S(p) = n, \forall p \in \Lambda$$

Then there is a closed neighborhood $N_\Lambda$ of $\Lambda$ such that for any two points $q_1$ and $q_2$ contained in $N_\Lambda$, there exists a piecewise control $u(t)$ such that the corresponding control orbit

$\phi(t,p,u))$ with $\phi(0,q_1,u) = q_1$ and $\phi(T,q_1,u)) = q_2$ for some $T > 0$.

*Proof* For each periodic orbit $\gamma \subset \Lambda$, it is to see from **Proposition 4.1** that there is an open neighborhood $N(\gamma)$ of $\gamma$ with the targetability property because of rank condition of $S$ on $\Lambda$. Since periodic orbits are dense in $\Lambda$, we have $\Lambda \subset \bigcup_{\gamma \in per(\phi)} N(\gamma)$, where $per(\phi)$ denotes the set of periodic orbits of $\Sigma_0$.

Since $\Lambda$ is compact, there are finite number of periodic orbits, say, $\gamma_1,\gamma_2,\cdots,\gamma_m$ in $\Lambda$, such that

$$\Lambda \subset \tilde{N} = \bigcup_{i=1}^{m} N(\gamma_i)$$

$\tilde{N}$ is clearly an open neighborhood of $\Lambda$.

It is easy to see that there is a closed (compact) neighborhood $N_\Lambda$ of $\Lambda$ satisfies $N_\Lambda \subset \tilde{N}$

Because $\Lambda$ is topologically transitive we can choose $N_\Lambda$ to be connected.

It remains to show that for any two points $p$ and $q$ in $\tilde{N}$, there is a piecewise control $u(t)$ such that the corresponding control orbit $\phi(t,p,u))$ with $\phi(0,p,u))=p$ and $\phi(T,p,u))=q$ for some $T>0$.

In fact, let $l$ be a path in $\tilde{N}$ that connects $p$ and $q$. Clearly $l$ can be regarded as a compact metric subspace in the states space which is covered by $\tilde{N}$. Then by Lebesgue lemma there are finite number of points $q_1,\cdots,q_s$ contained in $l$ with $q_1=p$ and $q_s=q$, such that every pair of points $\{q_i,q_{i+1}\}$ is contained in $N(\gamma_i)$ for some $1\le i\le m$.

By **Proposition 4.1,** for each pair $\{q_i,q_{i+1}\}$, there is a piecewise control $u_i(t)$ such that the corresponding control orbit has $q_i$ as its initial point at $t=0$ and $q_{i+1}$ as its final point at some $t_i>0$.

Now let
$$u(t)=\begin{cases}u_1(t), & t\in[0,t_1]\\ u_2(t), & t\in[t_1,t_1+t_2]\\ \cdots\cdots\\ u_{s-1}(t), & t\in[\sum_{i=1}^{s-2}t_i,\sum_{i=1}^{s-1}t_i]\end{cases}$$

The corresponding control orbit $t,\phi(p,u)$ connects $p$ and $q$. □

It is easy to see from the proof of the above theorem the statement still holds under weaker conditions.

**Theorem 4.2** Suppose $\Lambda$ is a chaotic set (attractor) of $\Sigma_0$. If for each periodic orbit $\gamma\subset\Lambda$, there is a point $p\in\gamma$, such that

$$\dim S(p)=n$$

Then the system is targetable in a neighborhood of $\Lambda$.

*Remarks* From the proof of **Theorem 4.1** it is easy to see that to guarantee the targetability of chaotic set, it is enough to find a finite number of periodic orbits that contain points satisfying rank condition in **Theorem 4.1**.

In control of chaos, one is more interested in finding a small perturbations (control) that just turn on in some small regions in state space, i.e., the targeting of chaotic attractors by small perturbations as called in the literature. To further develop our arguments on this problem we need a lemma that is also of interest of its own..

Let $B(\delta, p) = \{x \in R^n : \|x - p\| < \delta$,

**Lemma 4.1** Let $\gamma$ be a periodic orbit of $\Sigma_0$. Suppose that there exists a point $p \in \gamma$ such that

$$\dim S(p) = n$$

Then for any $\varepsilon > 0$ and $\delta > 0$, there is a neighborhood $N_\delta^\varepsilon$ of $\gamma$ such that $\Sigma$ is targetable in $N_\delta^\varepsilon$ and the control $u : [0, T] \to R^m$ satisfies

$$|u| \leq \varepsilon, \text{ and } u = 0 \text{ if } \phi(t, p, u)) \notin B(p, \delta).$$

where for $u = (u_1, ..., u_m) \in R^m$ one defines $|u| = \max\{|u_i|, i = 1, ..., m\}$

***Proof*** Let $U(\varepsilon) = \{u \in R^m : |u| \leq \varepsilon\}$. Since $U(\varepsilon)$ is bounded and $\phi(t, p, u)$ is continuous with respect to $t$ and $u$, there exists $T(\varepsilon, \delta) > 0$ such that for every small (piecewise) control

$$u : [0, T(\varepsilon, \delta)] \to U(\varepsilon)$$

We have

$$\phi(t, p, u) \in B(p, \delta), \quad t \in [0, T(\varepsilon, \delta)]$$

Thus

$$R(p, T(\varepsilon)) \subset B(p, \delta)$$

By **Proposition 2.1** the condition $\dim S(p) = n$ implies that

$$x(T(\varepsilon), p) \in \text{int } R(p, T(\varepsilon)) \subset B(p, \delta)$$

where $x(t, p)$ is the periodic orbit $\gamma$ of $\Sigma_0$ with initial point $x(0, p) = p$.

Let

$$r = \sup\{\lambda > 0 : B(p, \lambda) \subset \text{int } R(p, T(\varepsilon))\}$$

Then

$$B(p, r) \subset B(p, \delta).$$

Similarly from **Proposition 2.2** there exists $\overline{T}(\varepsilon, \delta) > 0$ such that for every small (piecewise) control

$$u : [-\overline{T}(\varepsilon, \delta), 0] \to U(\varepsilon)$$

We have

$$\phi(t,p,u) \in B(p,\delta), \quad t \in [-\bar{T}(\varepsilon,\delta),0]$$

Thus

$$R(p,-\bar{T}(\varepsilon,\delta)) \subset B(p,\delta)$$

Let

$$\bar{r} = \sup\{\lambda > 0 : B(p,\lambda) \subset \text{int } R(p,-\bar{T}(\varepsilon,\delta))\}$$

Then

$$B(p,\bar{r}) \subset B(p,\delta)$$

Denote by $\varsigma$ the period of $\gamma$ and define two sets

$$K^+ = \bigcup_{0 \leq t \leq \varsigma} x(t, B(p,r)) \quad \text{and} \quad K^- = \bigcup_{-\varsigma \leq t \leq 0} x(t, B(p,\bar{r}))$$

Then $K^+ \cap K^-$ is an open neighborhood of the periodic orbit $\gamma$ and clearly contains a neighborhood $N_\delta^\varepsilon$ of $\gamma$ such that $\Sigma$ is targetable in $N_\delta^\varepsilon$ and the control $u:[0,T(\varepsilon,\delta)] \to U(\varepsilon)$ can satisfy $u = 0$ if $\phi(t,p,u)) \notin B(p,\delta)$. $\square$

By means of this lemma we can easily prove the following main result in this paper.

**Theorem 4.3** Suppose that for each periodic orbit $\gamma \subset \Lambda$ there exists a point $p^\gamma \in \gamma$ such that

$$\dim S(p^\gamma) = n$$

Then for any $\varepsilon > 0$ and $\delta > 0$, there exists a neighborhood of $N_\Lambda$ of $\Lambda$, such that for any two points $q_1$ and $q_2$ contained in $N_\Lambda$, there exists a piecewise control $u(t)$ such that the corresponding control orbit $\phi(t,p,u))$ with $\phi(0,q_1,u) = q_1$ satisfies $\phi(T,q_1u)) = q_2$ for some $T > 0$, and the control $u(t)$ satisfies $|u(t)| < \varepsilon$. In addition there are finite number periodic orbits $\gamma_1, \gamma_2, \cdots, \gamma_m$ such that

$$|u(t)| = 0$$

if

$$\phi(t,p,u)) \in (R^n - \bigcup_{i=1}^m B(p^{\gamma_i},\delta)).$$

This means that if one wants to steer the obits in the neighborhood of $N_\Lambda$ of $\Lambda$, it is enough to consider the proper small control $u(t)$ in the small regions $B(p^{\gamma_i}, \delta)$, $i = 1, 2, ..., m$.

## 5 Conclusion

The so-called targeting in chaos, i.e. finding some suitable small perturbations (control strategies) to some system parameters in order to direct orbits contained in chaotic set to some desired final state is an important topic in the chaos control literature. Although a great deal of papers on this issue have been published up to now, the targetability of chaotic sets, or in terms of control theory, controllability of chaotic set by small system perturbations (control) has been seldom studied. This paper has presented a treatment on targetability of chaotic sets with small controls, and offered an affirmative by virtue of some results of geometric control theory.

**Acknowledgements** This work is supported in part by National Natural Science Foundation of China（10972082）.

Physica D 117, 1–12.